\documentclass[]{article}
\usepackage[applemac]{inputenc}	
\usepackage[english]{babel}
\usepackage{graphicx}
\usepackage{epsfig}
\usepackage{epstopdf}
\usepackage{psfrag}
\usepackage{subfigure}
\usepackage{amsmath}
\usepackage{bm}
\usepackage{amssymb}
\usepackage{siunitx}
\usepackage{upgreek}
\usepackage{color}
\usepackage{xcolor}
\usepackage{comment}
\usepackage{subeqnarray}
\usepackage{natbib}
\usepackage{xspace}
\usepackage{authblk}
\usepackage[margin=1.5in]{geometry}
\title{Entrainment of particles during the withdrawal of a fiber from a dilute suspension}
\author[]{B. M. Dincau$^1$, E. Mai$^{1,2}$,\,Q. Magdelaine$^{3,4}$, \, J. A. Lee$^{5}$\,,\, M. Z. Bazant$^{5,6,7}$, \& A. Sauret$^{1}$\thanks{Corresponding author: \texttt{asauret@ucsb.edu}}}
\affil{\small $^1$ Department of Mechanical Engineering, University of California, Santa Barbara, CA, USA \\
$^2$ Department of Chemical Engineering, University of California, Santa Barbara, CA, USA \\
$^3$ Surface du Verre et Interfaces, UMR 125 CNRS/Saint-Gobain, Aubervilliers, France \\
$^4$ Sorbonne Universit\'e, CNRS, Institut Jean Le Rond d'Alembert, F-75005 Paris, France \\
$^5$ Saint-Gobain Research North America, Northborough, MA, USA \\
$^6$ Department of Chemical Engineering, MIT, Cambridge, MA, USA \\
$^7$ Department of Mathematics, MIT, Cambridge, MA, USA \\}

\date{\small \today}

\begin{document}
\maketitle

\vspace{-5mm}
\begin{abstract}
A fiber withdrawn from a bath of a dilute particulate suspension exhibits different coating regimes depending on the physical properties of the fluid, the withdrawal speed, the particle sizes, and the radius of the fiber. Our experiments indicate that only the liquid without particles is entrained for thin coating films. Beyond a threshold capillary number, the fiber is coated by a liquid film with entrained particles. We systematically characterize the role of the capillary number, the particle size, and the fiber radius on the threshold speed for particle entrainment. We discuss the boundary between these two regimes and show that the thickness of the liquid film at the stagnation point controls the entrainment process. The radius of the fiber provides a new degree of control in capillary filtering, allowing greater control over the size of the particles entrained in the film.
\end{abstract}

\bigskip

\section{Introduction} \label{sec:intro}

Many industrial processes involve coating a substrate with a liquid film \cite[][]{scriven1988physics,weinstein2004coating}. Dip coating is a common method, where a substrate is withdrawn from a liquid bath at a controlled velocity $U$ to deposit a thin layer of liquid of thickness $h$ \cite[][]{quere1999fluid,rio2017withdrawing}. For a Newtonian liquid of dynamic viscosity $\mu$, density $\rho$ and surface tension $\gamma$, the entrainment of the liquid film is governed by the balance between viscous and capillary forces, whose ratio is measured by the capillary number ${\rm Ca} = \mu\,U / \gamma$. In the 2D situation, where a plate is withdrawn from a liquid bath, \cite{landau1942physicochim} and \cite{Deryagin} derived an expression for the film thickness for capillary number ${\rm Ca} < 10^{-2}$, $h=0.94\,\ell_c\,{\rm Ca}^{2/3}$, where $h$ is the thickness of the coating film and $\ell_c=\sqrt{\gamma/(\rho\,g)}$ is the capillary length.

Although most studies of dip coating have considered homogeneous fluids, various processes involve thin-film of suspensions containing particles. In industrial processes, this situation is involved in coating slurries on fibers for manufacturing of particle coated fiber additives and fabrics \cite[][]{gu2000deposition,wang2003light,jost2011carbon,wu2016preparation}, \textit{e.g.} containing optical coatings or ``smart'' particles or devices, but also for functional or smart fabrics and flexible detection/transmission devices, and as an alternate approach to achieve ``multi-material fibers'' \cite[][]{tao2015multimaterial}. Besides, dip coating can also be used for the filtration and sorting of particles \cite[][]{dincau2019capillary}. In some situations, the particle size becomes comparable to the thickness of the liquid film, and interfacial effects occur. While the effective viscosity $\mu(\phi)$ of a suspension, where $\phi$ denotes the volume fraction, can be obtained with classical rheological approaches \cite[][]{boyer2011,guazelli2011,guazzelli2018rheology}, interfacial effects can lead to unexpected observations, where the effective bulk rheology of the suspension fails to capture the flow dynamics. For instance, the presence of particles modify the pinch-off of suspension drops \cite[see \textit{e.g.}][]{furbank2004experimental,bonnoit2012}, the quasi-static break-up of a liquid bridge \cite[][]{McIlroy2014,lindner2015single,chateau2018pinch}, the stability of jets \cite[][]{hameed2009breakup,hoath2014jetted,chateau2019}, the fragmentation processes of particle-laden thin films \cite[][]{raux2020spreading}, but also the contact line dynamics \cite[][]{zhao2020spreading}. Besides, during the dip-coating of a plate the presence of particles leads to different coating regimes \cite[][]{Ghosh:2007ik,colosqui2013hydrodynamically,gans2019dip,palma2019dip}. For non-Brownian particles, at small capillary numbers corresponding to thin films, only the liquid is entrained while the particles remain trapped in the liquid bath. {When the withdrawal velocity satisfies $U>U^*$, where $U^*$ corresponds to the velocity threshold for individual particle entrainment, isolated particles are observed in the liquid film. This regime corresponds to capillary number Ca larger than $Ca^* \simeq 0.24 (a/\ell_c)^{3 / 4}$, for a plate, where $a$ denotes the particle radius.} For intermediate capillary numbers, {which depends on the volume fraction of the suspension and the withdrawal length}, clusters form in the meniscus before being entrained on the plate \cite[][]{sauret2019capillary}. {A fundamental feature of dip coating flows is the presence of a stagnation point $S^*$, which separates the flow in two regions: a region that continues into the coating film, and a recirculation flow region into the liquid bath [see Fig. \ref{fig:setup}(a)].} On a flat plate, the thickness of the liquid film at the stagnation point of the dynamic meniscus, $h^*$, controls the entrainment of wetted particles \cite[][]{colosqui2013hydrodynamically}. Furthermore, recent experiments on a 2D plate have demonstrated that individual particles are entrained if the radius of the particle $a$ satisfies the condition $\alpha\,a \leq h^*$, where $\alpha$ is a prefactor ($\alpha \simeq 1.1 \pm 0.1$) accounts for the complex shape of the meniscus around the particle. The existence of the different regimes have led to a new method of capillary filtration \cite[][]{dincau2019capillary}. Yet, this filtration is limited by the threshold for entrainment, which can only be controlled through the capillary length and the capillary number for planar substrate, thus limiting the range of particle sizes that can be filtered.

In this paper, we demonstrate that this limitation can be overcome by modifying the geometry of the substrate. Indeed, dip-coating with fibers exhibits a new length scale: the radius of the fiber, $ R < \ell_c $. The relevant parameter in this situation thus becomes the Goucher number ${\rm Go}=R/\ell_c$, which compares the ratio of the vertical curvature, set by $\ell_c$, and the azimuthal curvature set by the radius of the fiber $R$ \cite[][]{goucher1922problem}. {Note that one could also use the Bond number Bo defined as ${\rm Bo}={\rm Go}^{1/2}$}. For $ Go \ll 1 $, the thickness of the film is no longer governed by the capillary length $ \ell_c $ but by the radius of the fiber only \cite[][]{white1966theory,quere1999fluid}. We aim to describe how the particle entrainment threshold is modified by the cylindrical geometry. We consider experimentally a fiber withdrawn from a dilute particulate suspension. We begin by describing our experimental approach in Section 2. We then characterize the coating thickness in Section 3, where we highlight the role of the fiber geometry. Particle entrainment is characterized in Section 4, where we propose a theoretical explanation based on the thickness of the liquid film at the stagnation point, which controls the entrainment of particles in the coating film. Our theoretical model captures the experimental data and predicts the particle entrainment threshold for fiber substrates.


\section{Experimental methods} \label{sec:Experimental}

\begin{figure}
\centering
\subfigure[]{\includegraphics[width = 0.46\textwidth]{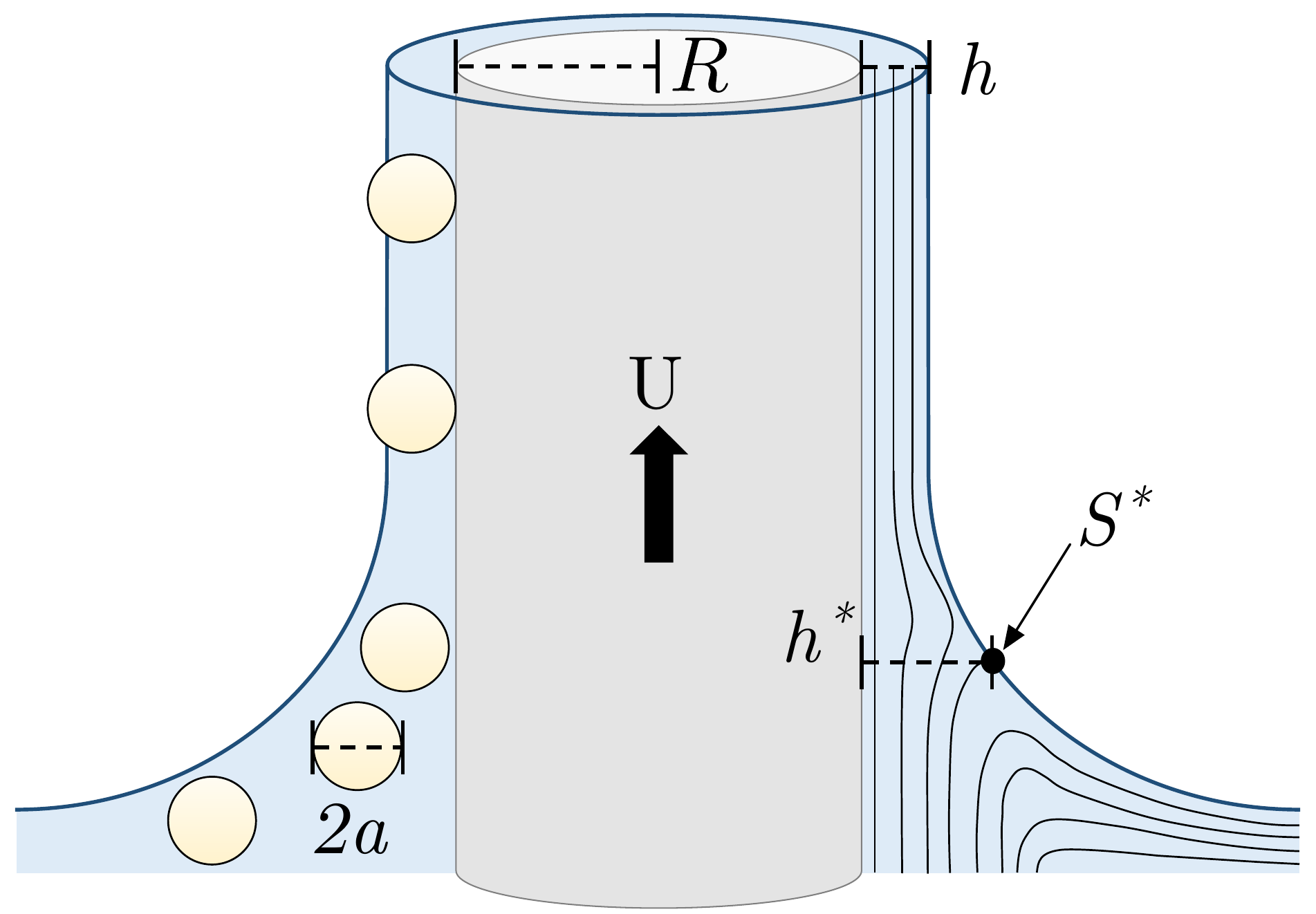}} \quad
\subfigure[]{\includegraphics[width = 0.46\textwidth]{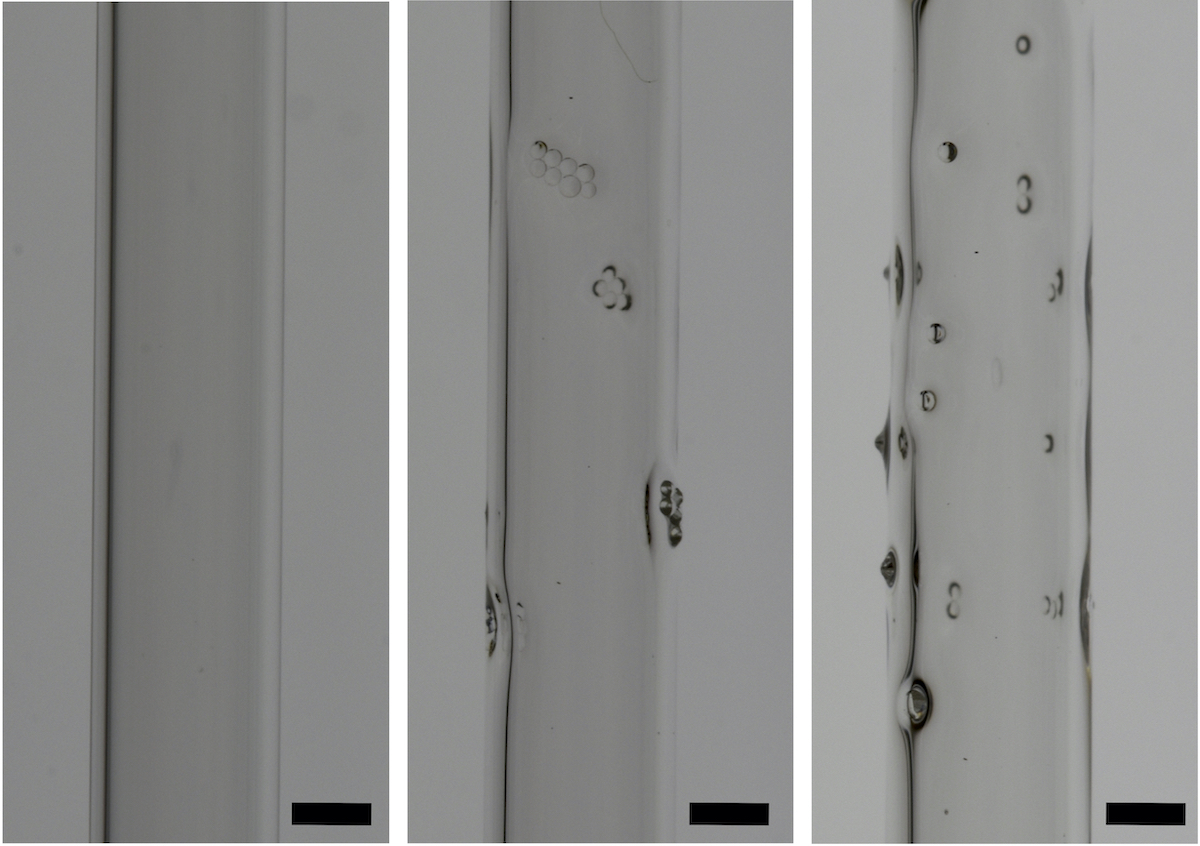}}
\caption{(a) Sketch of the experimental set-up. A fiber of radius $R$ is withdrawn from a suspension of particles of diameter $2\,a$ at a velocity $U$. (b) Examples of images obtained after the withdrawal of a fiber of radius $R=1.2\,{\rm mm}$ from a suspension of {$250\,\mu{\rm m}$} particles dispersed in AP 100. The velocity $U $ increases from left to right, which highlight three regimes: (i) no particle entrainment at low velocity, (ii) cluster entrainment just above the threshold, and (iii) mostly single particle entrainment well above the threshold. Scale bars are 1 mm.}
\label{fig:setup}
\end{figure}

The experimental setup is shown in Fig. \ref{fig:setup}(a). It consists of withdrawing a {glass} fiber {(Vitrocom)} of radius $R \in [125,\, 1200]\,\mu{\rm m}$ at a velocity $ U \in [0.01,\, 10]\,{\rm mm\,s^{-1}}$ from a bath of liquid containing a dilute suspension of spherical non-Brownian particles. The fiber is mounted in a fixed position to avoid any vibrations that would modify the film thickness, while a reservoir sits on a motorized linear translation stage {(NRT150/M with a BSC201 controller, Thorlabs)} beneath the fiber. The withdrawal dynamics are recorded using a digital camera {(Nikon D7200)} with a macro lens {(AF-S VR Micro-Nikkor $200 \, {\rm{mm}}$ f/4 lens)}. The fiber is back-lit with an LED panel {(Phlox)} mounted behind the container.

The suspensions are comprised of polystyrene particles (PS - Micro-Beads, Dynoseeds) of radius $a=[20,\,40,\,70,\,125,\,250]\,\mu{\rm m}$ and density $\rho_p=1056\,\pm 2\,{\rm kg\,m^{-3}}$. The particles are dispersed in a Newtonian fluid by mechanical stirring, ensuring a homogeneous suspension. We consider three different silicone oils : AP 100, AR 200 and AP 1000 (Sigma-Aldrich). {The physical properties of these fluids have been measured in a previous study \cite[][]{gans2019dip}}. {The different silicone oils are} of respective viscosity $\mu_{\rm AP100}=0.132\,{\rm Pa\cdot s}$, $\mu_{\rm AR200}=0.243 \,{\rm Pa\cdot s}$ and $\mu_{\rm AP1000}=1.42\,{\rm Pa\cdot s}$ ({measured with an Anton Paar MCR 501 rheometer and a plate/plate geometry}) and density $\rho_{\rm AP100}=1062\,{\rm kg.m^{-3}}$, $\rho_{\rm AR200}=1046\,{\rm kg.m^{-3}}$ and $\rho_{\rm AP1000}=1087\,{\rm kg.m^{-3}}$ ({measured with glass densimeters from VWR}), at $20^{\rm o}{\rm C}$. The interfacial tension of the silicone oils used in this study is $\gamma=21 \pm 1\,{\rm mN.m^{-1}}$ \cite{gans2019dip}. The volume fraction of the suspension, defined as the volume of particles $V_p$ divided by the total volume $V_{tot}=V_p+V_f$ is denoted $\phi =V_p/V_{tot}$ and remains smaller than $0.25\%$ in all the experiments such that the viscosity of the dilute suspension is equal to the viscosity of the interstitial fluid at first order {in the volume fraction $\phi$} and the particles can be considered as isolated. {Indeed, for volume fraction $\phi \leq 0.25\%$, the viscosity of the suspension can be obtained using the first linear correction in volume fraction to the viscosity $\mu(\phi)=\mu_0\,(1+5 \phi / 2)$, where $\mu_0$ is the viscosity of the interstitial fluid \cite[][]{guazzelli2018rheology}. In the present case, the maximum difference in viscosity is for the $250\,\mu{\rm m}$ particles and is equal to $0.63\%$.} {Between two experiments the suspension was thoroughly mixed and over the timescale of an experiment. The densities are matched within 3\% to reduce sedimentation. For all practical purposes the suspensions can be considered as neutrally buoyant as confirmed by our observation over the timescale of an experiment, typically between a few seconds to a few minutes \cite[see][for more details]{sauret2019capillary}.}

To determine the entrainment threshold of particles, we start the experiment with a low withdrawal velocity $U$, where no particles are entrained in the meniscus and only liquid coats the fiber, [left picture in Fig. \ref{fig:setup}(b)] while stirring in between each trial. We then increase the withdrawal speed incrementally until we observe the first particles coating the fibers [middle picture in Fig. \ref{fig:setup}(b)]. As we are working at a small volume fraction $\phi$, there is limited formation of clusters of particles. Further, from the velocity threshold $U^*$, we observe that the number of particles entrained increases with increasing the velocity [right picture in Fig. \ref{fig:setup}(b)]. The particle entrainment threshold is determined as the average of the last withdrawal speed where no particles are entrained and the first withdrawal speed where individual particles are entrained. The uncertainty on the estimate is the difference between these two velocities.


\section{Liquid film thickness}

\begin{figure}
	\centering
	\subfigure[]{\includegraphics[width = 0.495\textwidth]{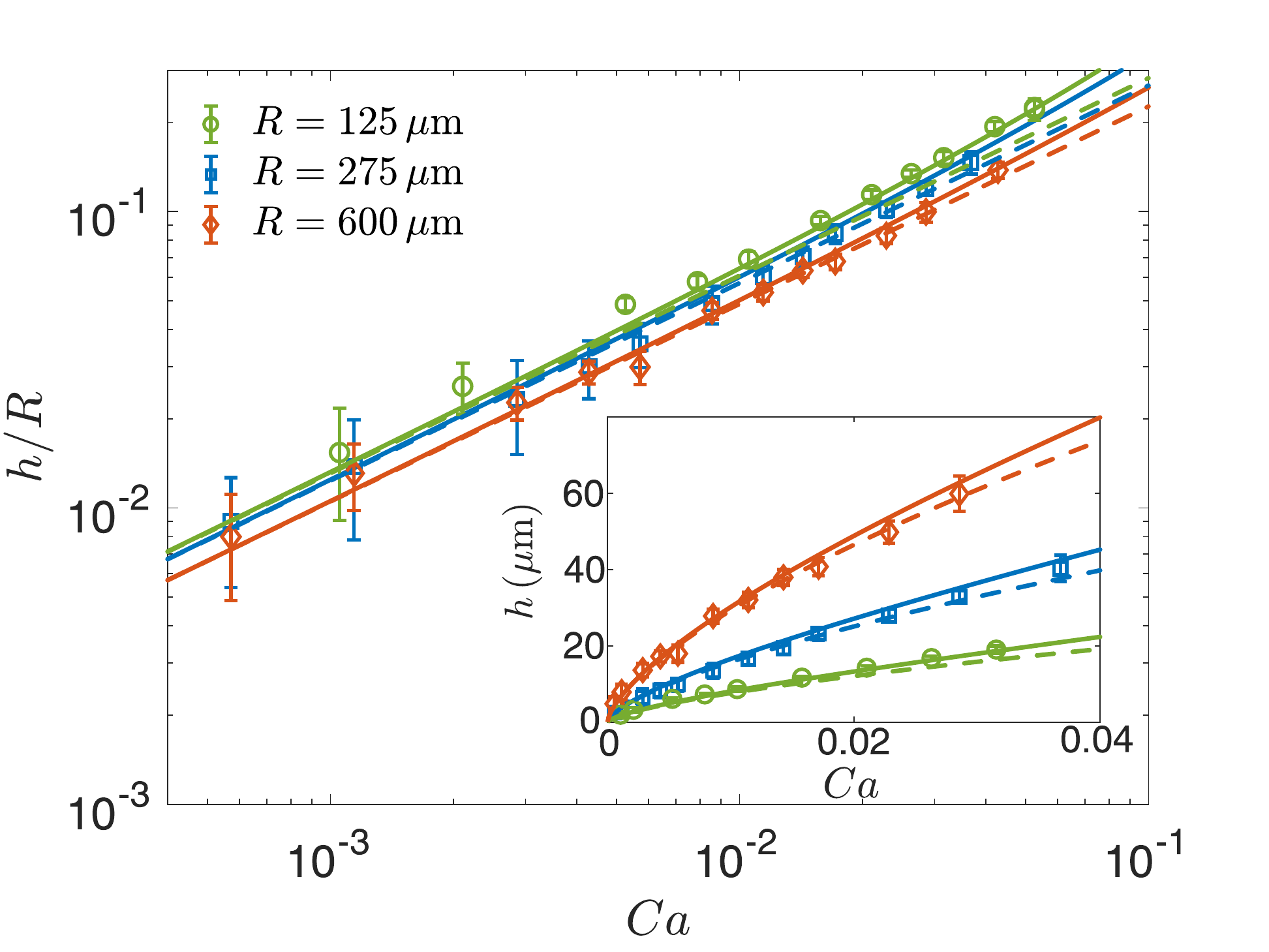}}
	\subfigure[]{\includegraphics[width = 0.495\textwidth]{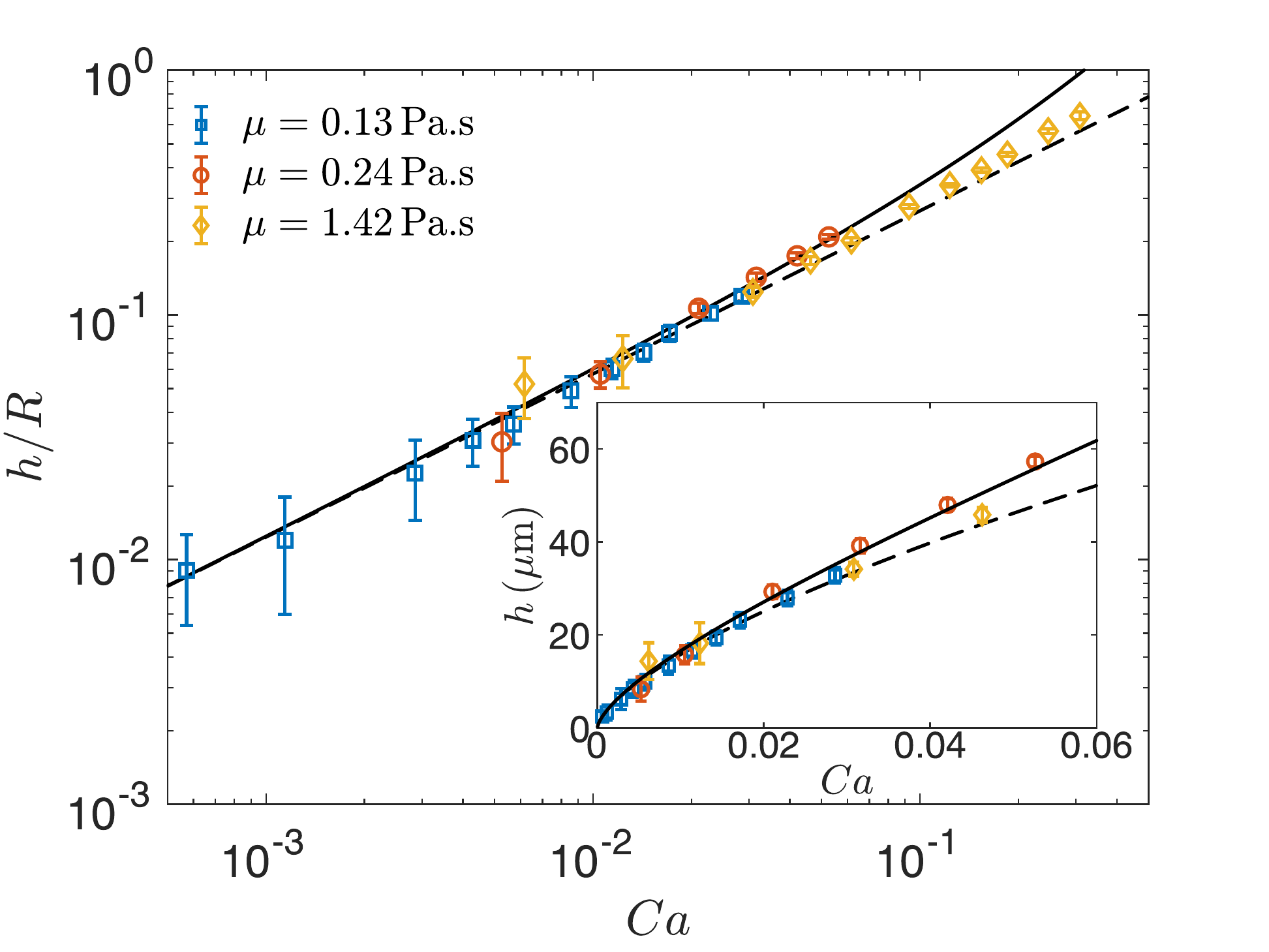}}
	\caption{Dimensionless liquid film thickness $h/R$ with (a) AP100 silicone oil ($\mu=0.132\,{\rm Pa\cdot s}$) and different fiber radii $R$ corresponding to ${\rm Go}=0.09$, $0.19$ and $0.42$; (b) fiber radius $R=275\,\mu{\rm m}$ (${\rm Go}=0.19$) and different silicone oils. In both figures, the insets report the liquid film thickness $h$, the symbols are the experimental measurements, the solid lines are the corresponding theoretical predictions given by the implicit system of equations \ref{system1}(a)-(b), and the dashed lines show the explicit expression provided by Eq. (\ref{Alex_thickness}).}
	\label{fig:liquid_only}
\end{figure}

Before considering the entrainment of particles, we first analyze the particle-free liquid film thickness entrained on withdrawn fibers. The thickness of the film coating a fiber withdrawn from a liquid bath has been less considered than the 2D configuration, \textit{i.e.} a plate \cite[][]{rio2017withdrawing}. The main difference associated with the cylindrical geometry is that the film thickness now depends on the ratio between the fiber radius $R$ and the capillary length $\ell_c$, captured through the Goucher number ${\rm Go}=R/\ell_c$. We performed systematic experiments measuring the entrained film thickness versus withdrawal speed varying the fiber radius (Fig. \ref{fig:liquid_only}a) and viscosity (Fig. \ref{fig:liquid_only}b). {Thanks to the cylindrical geometry, the film thickness can be determined by direct visualization of the edge of the fiber coated by the liquid. The processing is performed with ImageJ and a custom-made Matlab code.}

Fig. \ref{fig:liquid_only}(a) shows that the film thickness increases both with the capillary number and the fiber radius (fiber radii correspond to ${\rm Go}=0.09$, $0.19$ and $0.42$). Both Fig. \ref{fig:liquid_only}(a) and (b) show that the the capillary number dependence is captured by the well-known power law $h/R\propto{\rm Ca}^{2/3}$. {The fiber radius $R$ on the left hand side of this expression approximates the meniscus curvature far from the fiber where viscous effects become negligible, and is more generally expressed in terms of the ìstatic matchingî curvature $\kappa_{\infty}$. \cite{landau1942physicochim} first showed that by matching the flowing film curvature, in the lubrication approximation, to the limiting static curvature $\kappa_\infty $ results in the well-known relation
\begin{equation} \label{LL_general}
h \,\kappa_{\infty}=1.34 \,\mathrm{{\rm Ca}}^{2 / 3}.
\end{equation}
More specifically, \cite{landau1942physicochim} proposed to solve the meniscus drag out problem by breaking the meniscus into 3 regions: fully developed, transition, and static. In the fully developed region they argued that the film profile is exponential (solving equations in the appropriate limit), so the curvature is also exponential. In the static region they argued the curvature is that of a static meniscus, and approximated it to be constant, i.e., the curvature is parabolic. In the transition region, they numerically integrated the lubrication equations from the exponential boundary condition at the fully developed boundary up to ìfar enough awayî to match the constant condition. Thus, the matching curvature is the presumed curvature of the meniscus when it is nearly static. \cite{bretherton1961motion} later showed the same result in cylindrical geometry where the static curvature is set by the radius, in the context of long slender bubbles in tubes. For fiber withdrawal and entrainment, the approximation $\kappa_\infty=1/R$ is valid in the limits of very thin fibers, where the azimuthal curvature dominates over the planar curvature, ${\rm Go}<<1$, and thin films where the film surface radius is small compared to fiber radius, $h<<R$. This last condition is sometimes less precisely expressed as ${\rm Ca}<<1$, and thus for vanishingly thin films entrained on vanishingly small fibers the coating thickness is $h/R \approx 1.34 {\rm Ca}^{2/3}$ \cite[][]{quere1999fluid}.}

\cite{white1966theory} attempted to account for the planar curvature and non-vanishing film thickness, ultimately developing a semi-empirical expression for the matching curvature
\begin{equation} \label{Alex_1}
\kappa_{\infty} \approx \frac{1}{\ell_{c}}\left[\frac{\sqrt{2}}{2\, {\rm Go}\, s_{0}}+\frac{1.79\,\left({\rm Go}\,s_{0}\right)^{0.85}}{1+1.79\left({\rm Go} \,s_{0}\right)^{0.85}}\right]
\end{equation}
where $s_0=1+h/R$. {Note that the limit $s_0 \to 1 $ corresponds to $h\ll R$}. In the limit of infinite radius ${\rm Go}\rightarrow\infty$, the expression clearly reduces to the flat plate curvature $\kappa_{\infty}\rightarrow 1/\ell_c$. In the opposite limit, the curvature expression reduces not to the fiber radius itself, but the radius of the fully developed entrained film $h+R$; the further limit of thin films ($s_0\rightarrow 1$) clearly leads to $\kappa_{\infty}\rightarrow 1/R$.

Eq. \ref{Alex_1} and Eq. \ref{LL_general} together result in the implicit relation between the dimensionless film thickness and capillary and Goucher numbers:
\begin{subeqnarray} \label{system1}
	\frac{h}{R}=1+\frac{T\,{\rm Ca}^{1 / 2}}{{\rm Go}}, \\
	{\rm and} \quad T=0.944 \,{{\rm Ca}}^{1 / 6}\,\left[\frac{1.79 \,{{\rm Go}}^{0.85}\left(h / R\right)^{0.85}}{1+1.79 \,{{\rm Go}}^{0.85}\,{\left(h / R\right)^{0.85}}}+\frac{0.71}{{{\rm Go}}\left(h / R\right)}\right]^{-1}.
\end{subeqnarray}
Resolution of this system leads to the solid curves in Fig. \ref{fig:liquid_only}, which compare favorably with all experiments {as it captures all the important trends: with respect to the capillary number Ca and with respect to the Goucher number Go}. {Fig. \ref{fig:liquid_only}(b) shows that the predictions of Eq. \ref{system1}(a)-(b) (solid curve) departs from the simple power law dependence (dashed line) at moderate values of capillary number, reflecting departure from the thin film limit. The small discrepancy in the film thickness could also be due to the drainage of the film as observed for 2D plate \cite[][]{maleki2011landau}.} For small capillary number and within uncertainty, the experiments do not exhibit a departure from the simple power law (dashed line) that Eqs. (\ref{system1}) exhibits, suggesting that some simpler limit may suffice to describe them. Retaining the finite Goucher number dependence but invoking the thin film limit ($s_0\rightarrow 1$) in Eq. \ref{Alex_1} results in the simplified matching curvature expression:
\begin{equation}
\kappa_{\infty} \approx \frac{\sqrt{2}}{\ell_{c}}\,\left[\frac{\sqrt{2}}{2\, {\rm Go}}+\frac{1.79\,\mathrm{Go}^{0.85}}{1+1.79\,\mathrm{Go}^{0.85}}\right].
\end{equation}
In this limit, Eq. \ref{system1} affords explicit expression of the dimensionless film thickness $h/R$ in terms of the capillary number and Goucher number:
\begin{equation} \label{Alex_thickness}
\frac{h}{R}=\frac{1.34 \,\mathrm{Ca}^{2 / 3}}{1+2.53\,\mathrm{Go}^{1.85}/[1+1.79\,\mathrm{Go}^{0.85}]}.
\end{equation}
In particular, Eq. \ref{Alex_thickness} retains the simple power law dependence on $\rm Ca$ while accounting for finite fiber radius. Its predictions are represented by the dashed curves in Fig. \ref{fig:liquid_only}, which appear to capture both of the relevant dependences in capillary and Goucher numbers while retaining a convenient simplicity. Therefore, the forthcoming analysis of particle entrainment will be based on Eq. \ref{Alex_thickness}.


\section{Dilute suspension: entrainment of isolated particles}

\subsection{Experimental results}

\begin{figure}
\centering
\includegraphics[width = 0.95\textwidth]{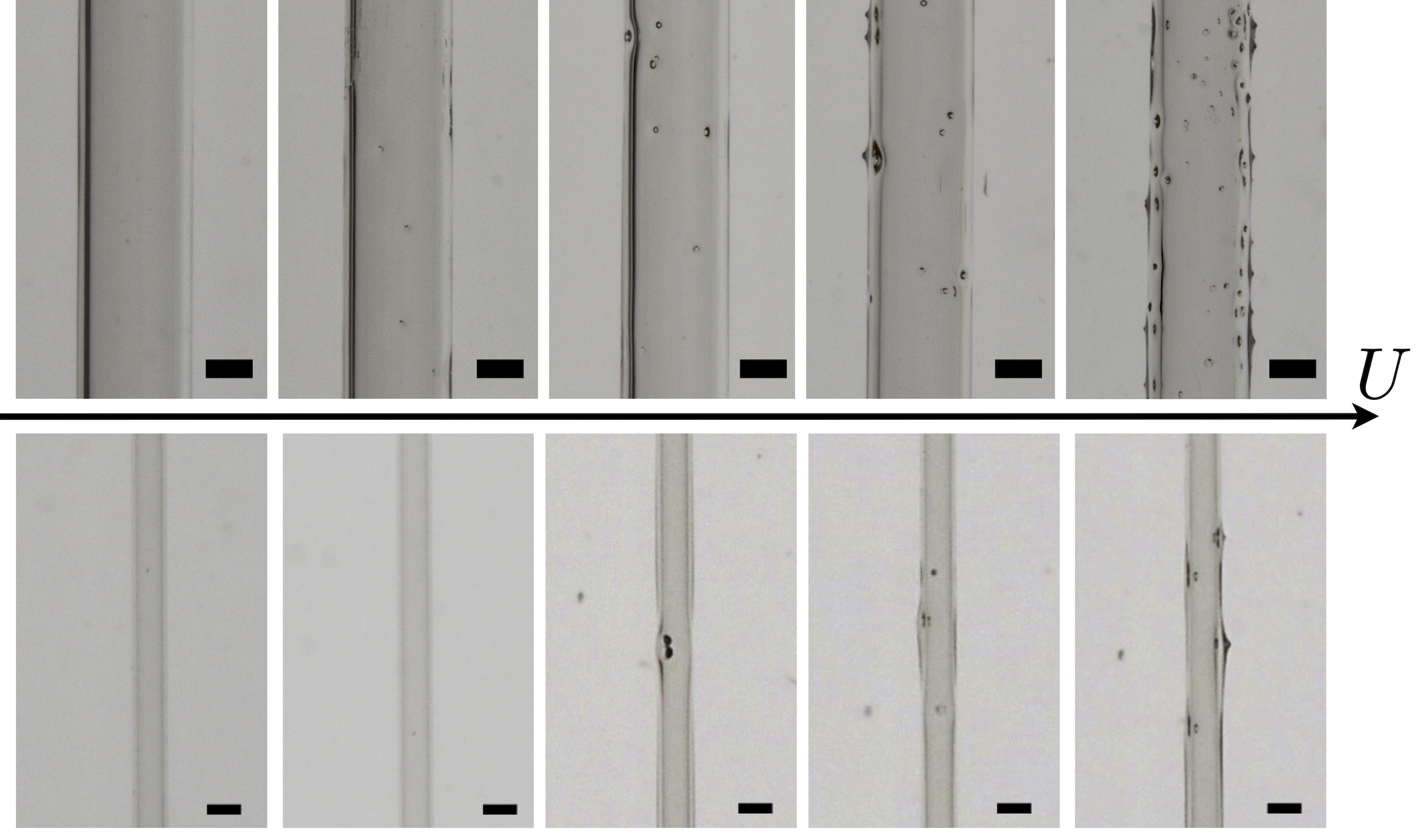}
\caption{Examples of fibers after their withdrawal from a suspension. Top pictures: $a=70\,\mu{\rm m}$ particles dispersed in AP 100, for a fiber of radius $R=1.2\,{\rm mm}$ and increasing velocities $U=0.5,\,1,\,3,\,5,\,7\,{\rm mm\,s^{-1}}$ (from left to right). Scale bars are 1 mm. Bottom pictures: $a=70\,\mu{\rm m}$ particles dispersed in AP 1000 for a fiber of radius $R=0.55\,{\rm  mm } $ and increasing velocities of  $U=1,\,2,\,3,\,4,\,5\,{\rm mm\,s^{-1}}$. Scale bars are $500\,\mu{\rm m}$.}
\label{fig:exp1}
\end{figure}

We perform a systematic study varying the particle radius $ a $, the fiber radius $ R $, and the viscosity of the interstitial fluid $ \mu $. {Here the capillary length is of the same order for all silicone oils considered and equal to $\ell_c \simeq 1.4\,{\rm mm}$.} Examples of experimental observations are reported in Fig. \ref{fig:exp1}. For a given fiber and particle radii, there is a threshold velocity $U^*$ below which only a liquid film coats the fiber, and the particles remain trapped in the liquid bath. Beyond this threshold, particles start to cover the fiber. {Therefore}, three regimes are observed, similar to the 2D situation with the withdrawal of a plate \cite[][]{sauret2019capillary}: a coating film (i) without particles as shown in the left picture in Fig. \ref{fig:setup}(b), (ii) with clusters of particles formed in the meniscus and entrained in the coating film, visible on the middle picture in Fig. \ref{fig:setup}(b) and (iii) with individual particles as reported in the right picture in Fig. \ref{fig:setup}(b). {The different regimes are clearly visible in Fig. \ref{fig:setup}(b) where the volume fraction is slightly higher that what we considered for the systematic measurements of the entrainment threshold. As shown by \cite{sauret2019capillary}}, the entrainment of clusters depends on the probability of forming a cluster of large enough size in the meniscus, whereas the entrainment of individual particles only requires that they be in close enough proximity to the fiber and overcome the resistive capillary force induced by the air-liquid meniscus. {Therefore, at very small volume fraction, the range of existence of the cluster regime is very limited and make it easier to estimate the entrainment threshold. For this reason, we consider only dilute suspensions in this work} making it easier to focus on the threshold velocity $U^*$ for isolated particle entrainment.

\begin{figure}
\centering
\subfigure[]{\includegraphics[width = 0.45\textwidth]{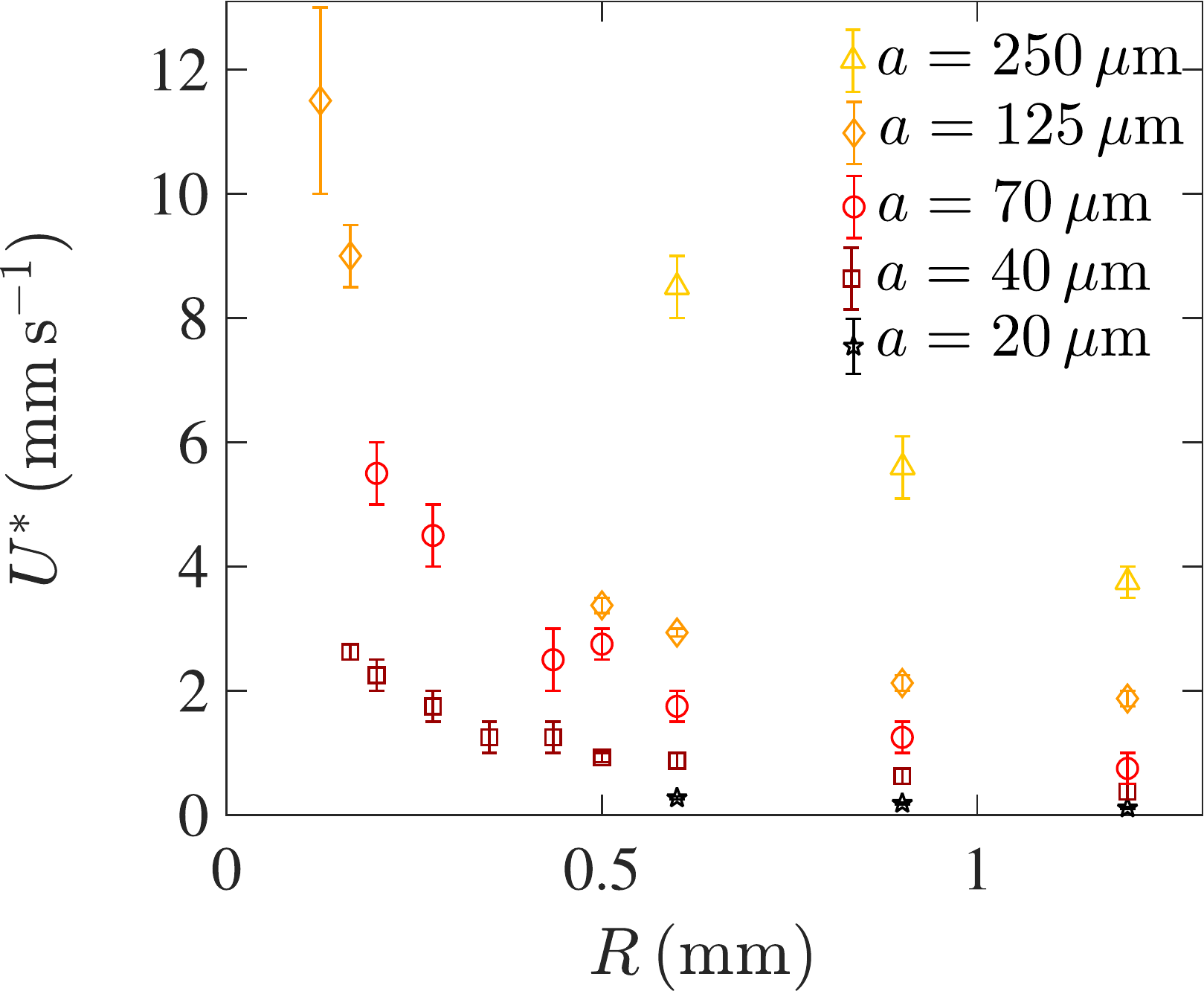}} \quad
\subfigure[]{\includegraphics[width = 0.45\textwidth]{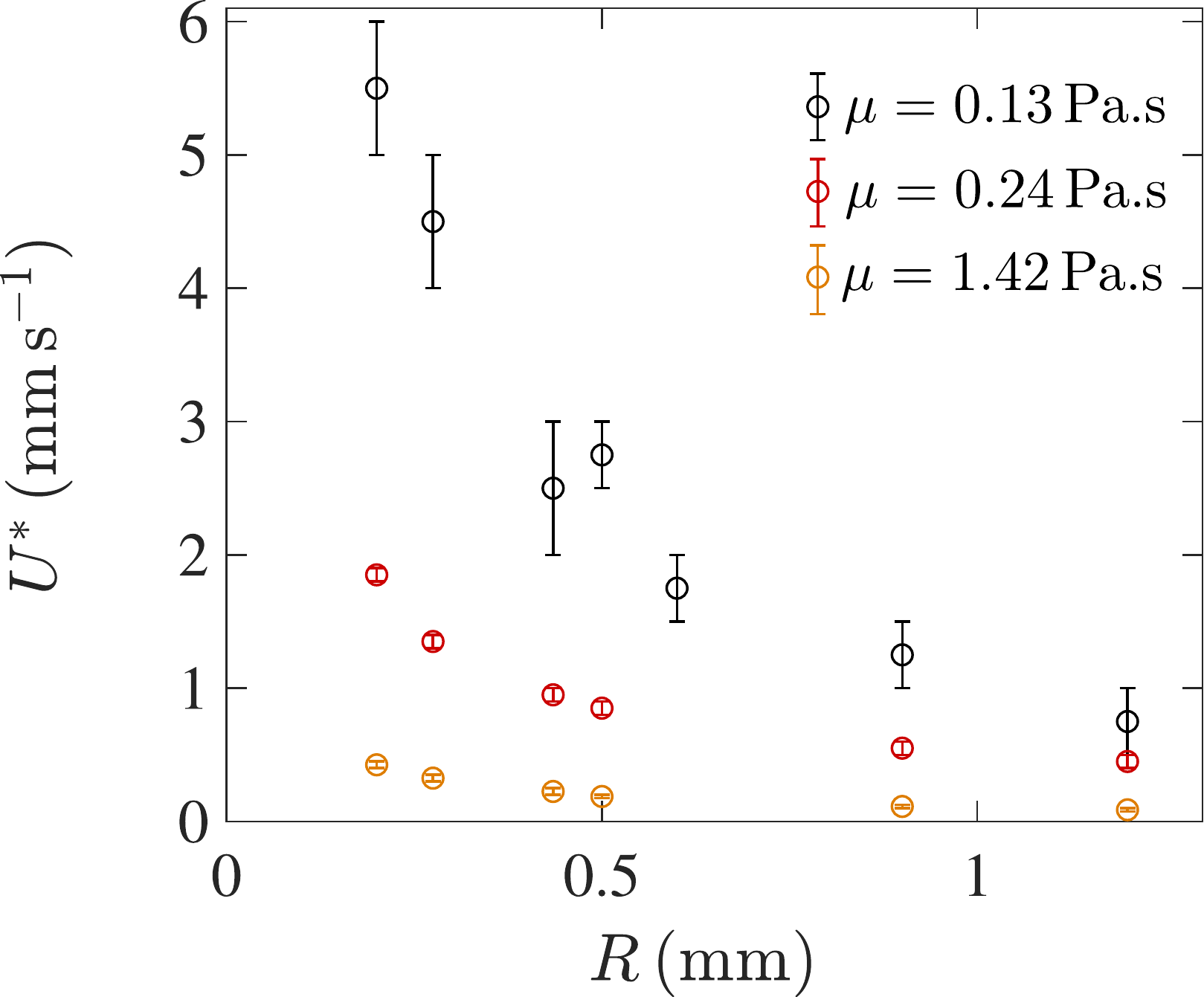}}
\caption{Withdrawal velocity threshold $U^*$ for particle entrainement in the coating film as a function of the radius of the fiber $R$ for various (a) particle radii $a$ dispersed in the silicone oil AP 100 and (b) different viscosity of the interstitial fluid for particles of radius $a=70\,\mu{\rm m}$ dispersed in silicone oils.}
\label{fig:exp_2}
\end{figure}

The entrainment threshold $U^*$ increases with the radius of the particles $a$ for a given fiber diameter and is not influenced by inertial effects. {Indeed, the maximum value of the Reynolds number, defined as ${\rm Re}=\rho\, h\, U / \mu$, is obtained for the thicker films corresponding to $\rho \simeq 1060\,{\rm kg.m^{-3}}$, $h\sim 50\,\mu{\rm m}$, $U \sim 10^{-2}\,{\rm m.s^{-1}}$ and $\mu=0.132\,{\rm Pa\cdot s}$, and satisfies $Re < 10^{-2}$. We can thus consider that we are always in a viscous regime where ${\rm Re} \ll 1$.} The threshold is directly related to the thickness of the coating fluid on the fiber, which increases with the speed of withdrawal. Therefore, the entrainment of larger particles on a fiber of radius $R$ requires a larger withdrawal speed. The experimental trend observed here for the influence of the particle size is similar to the trend reported for 2D plates \cite[][]{sauret2019capillary}.

The main difference brought by the cylindrical geometry is that the particle entrainment threshold not only depends on the size of the particles $a$ and the viscosity of the interstitial fluid $\mu$ but also on the radius of the fiber $R$. For fibers of different radii and the same interstitial fluid and particle size, Fig. \ref{fig:exp_2}(a) illustrates that the entrainment of particles does not occur at the same threshold velocity. The entrainment of particles in the liquid film occurs at a larger withdrawal speed for the smaller fibers. Therefore, the influence of the fiber radius provides a new degree of control of the entrainment threshold. {We should also emphasize that when the diameter of the particle $2\,a$ becomes of the same order than the radius of the fiber $R$, the entrainment of particles will be difficult to achieve. Indeed, a coating film of the order of the particle radius would require much higher capillary number in this regime and our model for the coating film thickness would not be valid anymore.}

The influence of the viscosity is similar to the 2D situation, as illustrated in Fig. \ref{fig:exp_2}(b). Higher fluid viscosity is associated with larger capillary numbers, ${\rm Ca}=\mu\,U/\gamma$. For a constant withdrawal velocity and a fixed surface tension, the coating film thickness thus increases with the viscosity, and the entrainment threshold increases when decreasing the viscosity of the fluid at a fixed $R$.

Our experiments demonstrate that the larger the size of the fiber, the smaller the particle entrainment threshold. This effect can only be explained by considering the dependence of the film thickness on the fiber radius. 

\subsection{Stagnation point} \label{sec:stagnation}

The thickness at the stagnation point, $h^*$ (see Fig. \ref{fig:setup}), has been derived in the 2D situation, \textit{i.e.}, for a plate exiting a liquid bath \cite[][]{colosqui2013hydrodynamically} but has not been reported in the fiber configuration to the best of our knowledge.

We consider the withdrawal of a fiber of radius $R \ll \ell_c$. We use the cylindrical coordinates $r$ and $z$ (radial and along the axis of the fiber) and denote $p$ the pressure and $\boldsymbol{u}=(u_r,\,u_\theta,\,u_z)$ the fluid velocity in these coordinates. In our experimental situation, we do not have surfactants, which could otherwise modify the interfacial tension and introduce Marangoni effects \cite[][]{shen2002fiber}. Zero normal velocity and zero normal and tangential stress balance boundary conditions need to be satisfied at the air-liquid interface, at $r=R+h(z)$. At the surface of the fiber, the no-slip boundary condition is $u_z(r=R)=0$. Dip coating dynamics on a substrate lead to thin film, we thus use the standard lubrication equations. The velocity in the direction normal to the fiber $u_r$ is negligible compared to the velocity along the fiber $u_z$. The pressure $p$ is obtained from the Laplace equation in the axisymmetric meniscus. We define the position of the free surface of the liquid film as $\xi(z)=R+h(z)$. The pressure in the thin film approximation (${\rm d}_z\xi \ll 1$) is $p=\gamma\left[{1}/{\xi}-{{\rm d}^2\xi}/{{\rm d} z^2}\right]$. Using the boundary conditions, we obtain the velocity field:
\begin{equation}\label{velocity}
\frac{u_z}{U}=1-\frac{1}{4\,{\rm Ca}}\left(\frac{{\rm d}^3\xi}{{\rm d} z^3}+\frac{1}{\xi^2}\,\frac{{\rm d}\xi}{{\rm d} z}\right)\left[r^2-R^2-2\,\xi^2\,{\rm ln}\left(\frac{r}{R}\right)\right].
\end{equation}

The corresponding flow rate $Q$ is constant and given by
\begin{equation}\label{Qgeneral}
\frac{Q}{U\,\pi\,R^2}=\left(\frac{\xi^2}{R^2}-1\right)-\frac{R^2}{{\rm Ca}}\left(\frac{{\rm d}^3\xi}{{\rm d} z^3}+\frac{1}{\xi^2}\,\frac{{\rm d}\xi}{{\rm d} z}\right)\,\mathcal{F}\left(\frac{\xi}{R}\right),
\end{equation}
where the function $\mathcal{F}(x)$ is defined as $\mathcal{F}(x)={1}/{8}+{3\,x^4}/{8}-{x^2}/{2}-{x^4}{\rm ln}\,(x)/2 $. The stagnation point $S^*$ corresponds to the location of the interface where the surface velocity vanishes. The expression of the surface velocity is obtained from Eq. (\ref{velocity}) evaluated at $r=\xi$. Since the thickness of the liquid film $h(z)$ is much smaller than the radius of the fiber, $h/R \ll 1$, the surface velocity $u_s$, corresponding to the velocity tangential at the interface, is given by
\begin{equation}\label{surface_velocity}
\frac{u_s(z)}{U}=1+\frac{h^2}{2\,{\rm Ca}}\left(\frac{{\rm d}^3h}{{\rm d} z^3}+\frac{1}{R^2}\,\frac{{\rm d}h}{{\rm d} z}\right).
\end{equation}
The second term of this equation can be simplified by using the flow rate far from the meniscus, $Q_{\infty}=2\,\pi\,U\,h\,R$. Using this expression with Eq. (\ref{Qgeneral}) allows us to simplify the expression (\ref{surface_velocity}) for the surface velocity:
\begin{equation}
\frac{u_s}{U}=1-\frac{h^2\left({\xi_\infty}^2-\xi^2\right)}{2\,R^4\,\mathcal{F}(\xi/R)}.
\end{equation}
Therefore, in the limit $h/R\ll 1$, the thickness at the stagnation point, \textit{i.e.}, where the surface velocity vanishes, is equal to ${h^*}={3}\,{h}$. The value obtained for a fiber is thus {the same as} the value obtained for a 2D plate {for $Ca\ll 1$} \cite[][]{colosqui2013hydrodynamically}.

\subsection{Numerical simulation: coating thickness and stagnation point} \label{sec:numerical}

The comparison of this calculation to experiments is challenging as it requires visualizing the streamlines in a very narrow region. We thus rely on numerical simulations to determine the position of the stagnation point. {The numerical simulations are performed with the Basilisk open-source library \footnote{www.basilisk.fr}, successor of Gerris \cite{popinet2009accurate}, using an adaptive mesh and a volume-of-fluid method to describe the two phases, liquid and air, and their interface. We solve the unsteady Stokes equation with a homogenous surface tension, assuming the problem to be axisymmetric and the fluids to be incompressible and Newtonian. The ratio of density and viscosity between the two phases are respectively fixed to 100 and 44.6. The surface tension and the gravity acceleration are implemented with the continuum surface force (CSF) method, as described for example by \cite{popinet2018numerical}. For this implementation of the gravity acceleration, the hydrostatic pressure is analytically subtracted to the mechanical pressure, leading to an effective pressure and reducing the gravity to a surface force. The numerical domain is a square, whose expansion is chosen to be around three times the capillary length. It includes part of the bath from which the fiber is redrawn. On the left side, we assume a no-slip condition of the fluids on the fiber, by imposing a homogeneous Dirichlet condition for the normal velocity and a Dirichlet condition equal to the velocity of the fiber for the tangential component. Free flow boundary conditions are assumed on the three other boundaries. Thanks to the reduced gravity formulation, these conditions are easily obtained, applying a homogeneous Dirichlet condition for the effective pressure and a homogeneous Neumann condition for the velocity field. The time step is determined by the CourantñFriedrichsñLewy condition conditions \cite[][]{popinet2018numerical}. The simulation is initialized with a thin film on fiber linked to the bath by a meniscus. The simulation is stopped after the fiber has browsed two or three times the height of the numerical domain.}

\begin{figure}
\centering
\subfigure[]{\includegraphics[width = 0.45\textwidth]{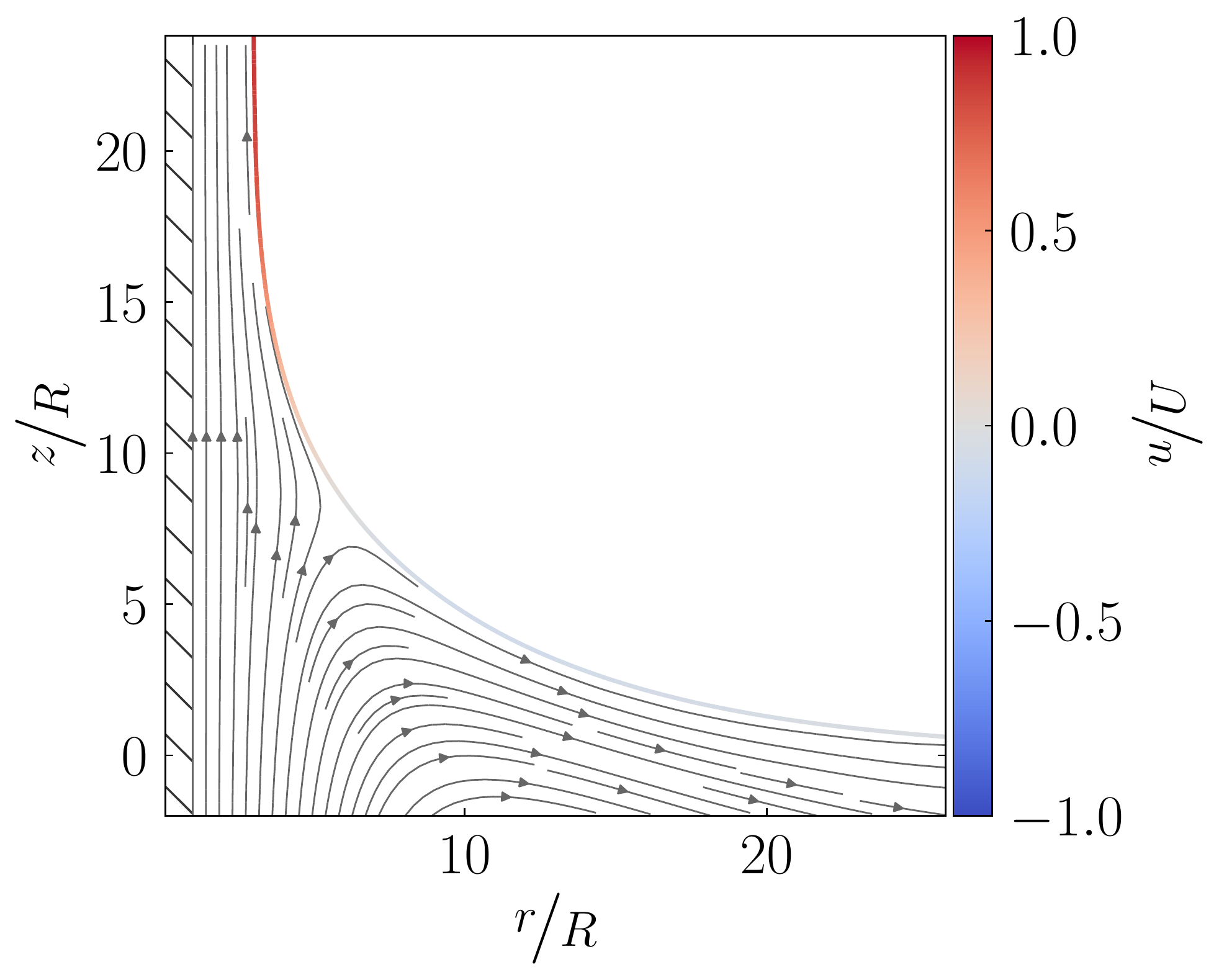}}
\subfigure[]{\includegraphics[width = 0.50\textwidth]{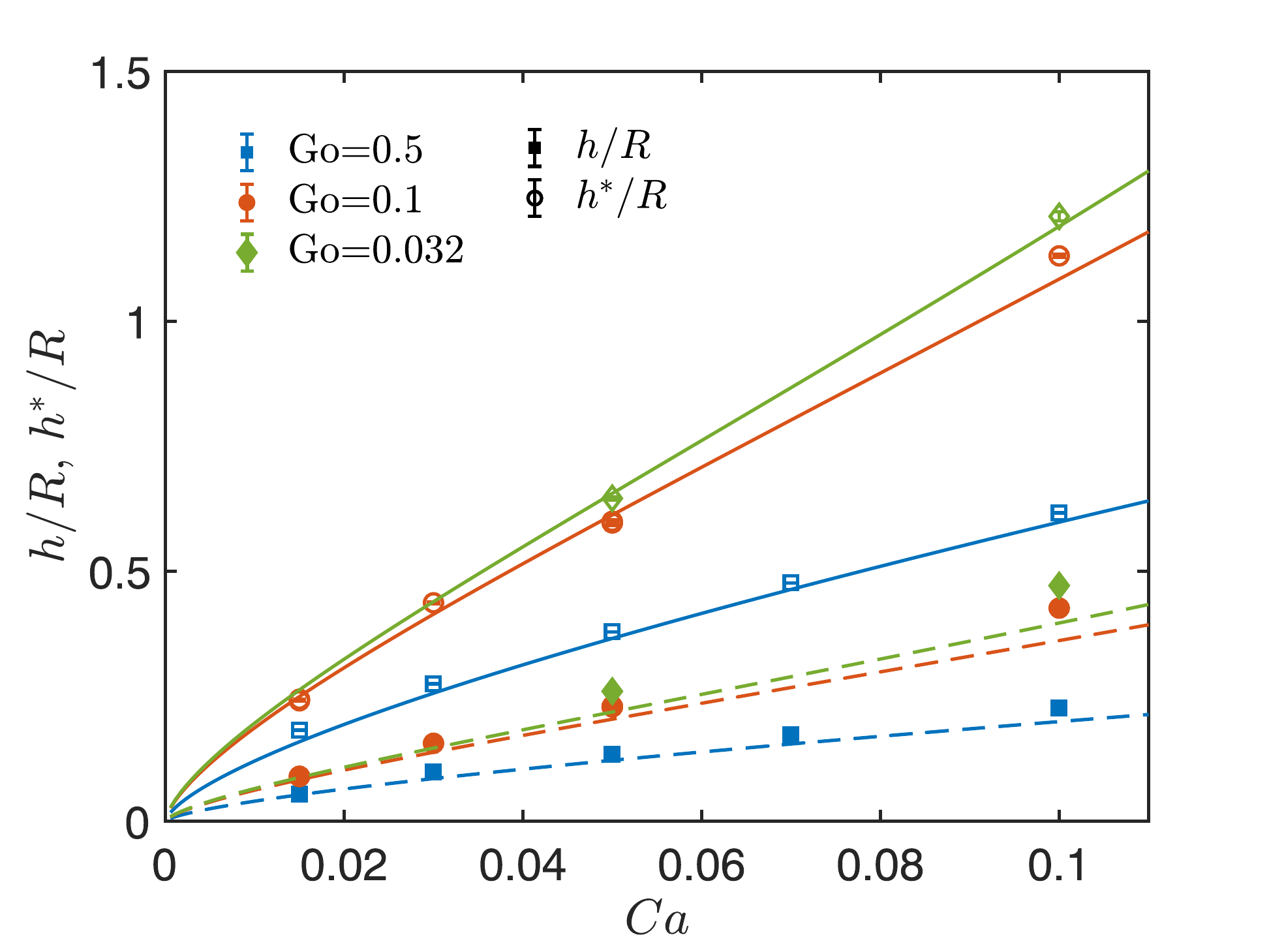}}
\caption{Numerical simulation of the dip coating of a fiber. (a) Examples of streamlines observed around the meniscus. The color bar indicates the dimensionless surface velocity. The stagnation point corresponds to the point where this surface velocity vanishes. (b) Evolution of the rescaled film thickness $h/R$ (filled symbols) and the rescaled thickness at the stagnation point $h^*/R$ (open synbols) for different values of the Goucher number ${\rm Go}$. The dashed lines correspond to the theoretical expressions of the thickness given by Eq. (\ref{Alex_thickness}) and the solid lines to the theoretical thickness at the stagnation point, $h^*/R=3\,h/R$.}
\label{fig:quentin}
\end{figure}

Numerically, the streamlines exhibit a clear stagnation point, as illustrated by the example in Fig. \ref{fig:quentin}(a). The stagnation point is obtained by estimating the location where the surface velocity vanishes. At this location, the streamlines define the transition between the fluid that continues into the coating film, and the liquid that recirculates in the liquid bath. We report the evolution of the film thickness far from the meniscus $h$ and the thickness at the stagnation point $h^*$ for fibers associated with different Goucher numbers, ${\rm Go}=0.032,\,0.1,\,0.5$, in Fig. \ref{fig:quentin}(b). The results of the numerical simulations agree well (within 10\%) with the theoretical prediction of the thickness at the stagnation point $h^*/R=3\,h/R$ for a capillary number smaller than ${\rm Ca} < 10^{-1}$.

\subsection{Threshold for particles entrainment}

For a particle to be entrained in the coating film, the thickness at the stagnation point $h^*$ must be larger than a fraction of the particle diameter $2\,a$, so that the viscous drag exerted on the particle overcomes the capillary resistance. {The presence of the particle of the same size as the thickness at the stagnation point $h^*$ will disturb the flow. Nevertheless, comparing $h^*$ and the size of the particle have been shown to predict satisfactorily well the entrainment threshold \cite[][]{colosqui2013hydrodynamically,sauret2019capillary} and a similar criterion is used here.} This criterion is expressed as:
\begin{equation}\label{eq:criterion}
\alpha\,a \leq h^*
\quad
{\rm where}
\quad
0<\alpha\leq 2.
\end{equation}
Experiments performed with spherical particles entrained on a 2D plate indicate that the prefactor $\alpha$ accounts for the complex shape of the meniscus around the particle, and reported $\alpha \simeq 1.1 \pm 0.1$. Such results are expected to hold for fibers in the limit of particle radii much smaller than the fiber radius, $a/R \ll 1$, since locally the particle sees the coating film as flat. But as the pure liquid film thickness analysis showed, the finite fiber radius must still be considered when estimating the fully developed film thickness around the fiber, and thus the thickness at the stagnation point.

{We have demonstrated analytically in section \ref{sec:stagnation} and confirmed through numerical simulation in section \ref{sec:numerical} that a sufficient estimate for the stagnation point thickness is $h^*=3\,h$}. From this expression of $h^*$, together with the criterion of Eq. (\ref{eq:criterion}) leads to a critical condition on the fully developed film thickness $h=\alpha a/3$. This film thickness can be inserted into either Eqs. (\ref{system1}) or Eq. (\ref{Alex_thickness}) to obtain a relationship between particle's size $a$ and the threshold capillary number ${\rm Ca}^*$ for entraining it.

The relationship according to the thin film limit (Eq. \ref{Alex_thickness}) is 
 \begin{equation}
\frac{\alpha\,a}{3\,R}=\frac{1.34 \mathrm{{\rm Ca^*}}^{2 / 3}}{1+2.53\,\mathrm{Go}^{1.85}/\left[1+1.79\,\mathrm{Go}^{0.85}\right]},
 \end{equation}
where $a/R$ is the ratio of the particle to fiber radius. Re-arranging for the capillary number leads to:
\begin{equation}\label{ref_finale}
 {\rm Ca}^* \simeq 0.645 \left[\frac{\alpha\,a}{3\,R}\left(1+\frac{2.53\,\mathrm{Go}^{1.85}}{1+1.79\,\mathrm{Go}^{0.85}}\right)\right]^{3/2},
\end{equation}
In the limit of vanishing Goucher number, the threshold capillary number follows a simple power law in the particle size:
\begin{equation}\label{ref_finale2}
{\rm Ca}^*=0.645\left[\frac{\alpha\,a}{3\,R}\right]^{3/2}.
\end{equation}
The Goucher number dependence and particle size depdence in Eq. (\ref{ref_finale}) are separable such that the threshold capillary number can be rescaled by the Goucher number dependence to follow the single power law on the right hand side of Eq. (\ref{ref_finale2}).

\begin{figure}
\centering
\includegraphics[width = \textwidth]{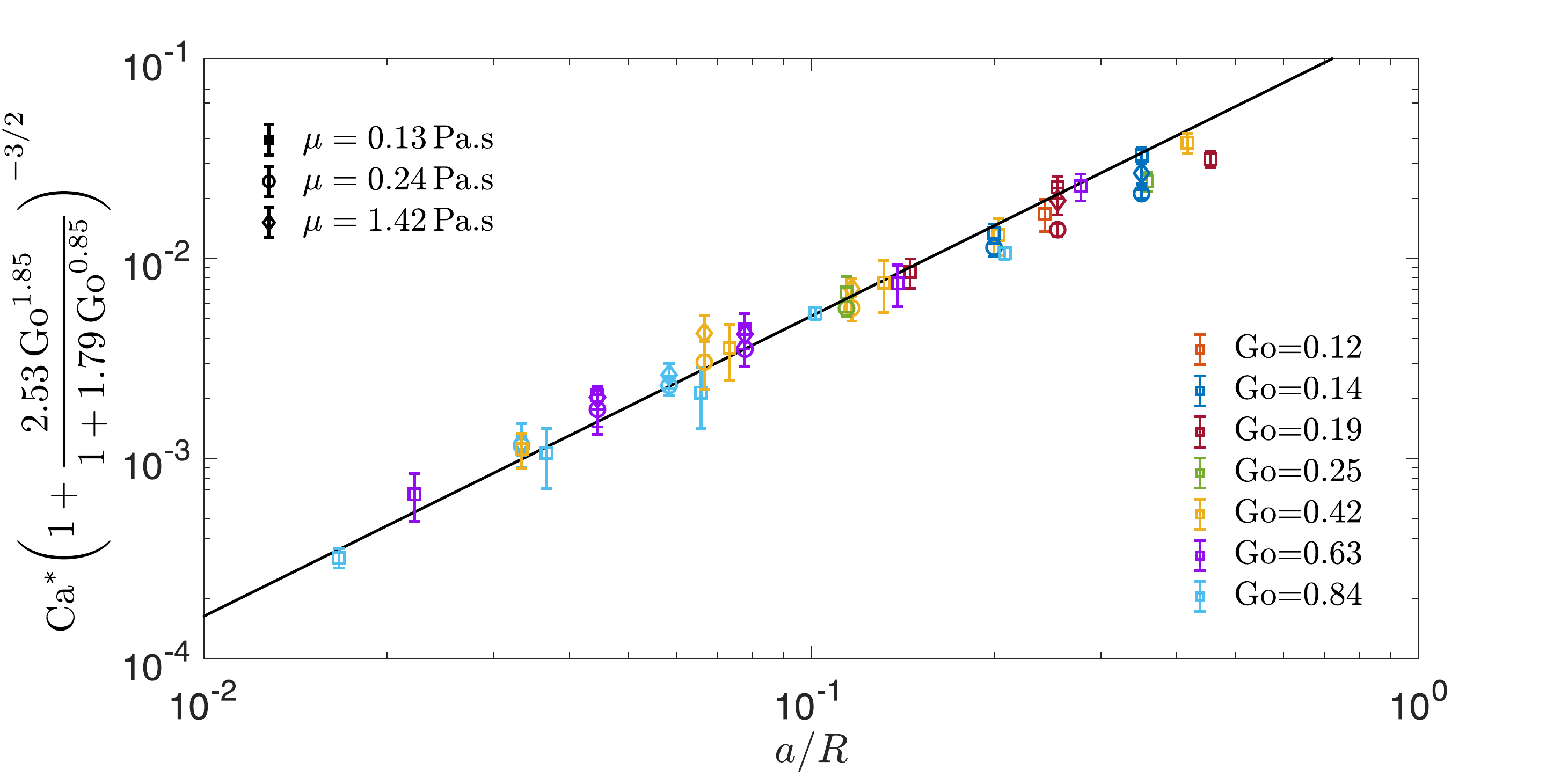}
\caption{Capillary number threshold ${{\rm Ca}}^*$ for particles entrainement as a function of the ratio $a/R$ for a fiber of various radii $R \in [165,\,1200]\,\mu{\rm m}$ leading to Goucher number ${\rm Go} \in [0.12,\,0.84]$. The solid line is the expression (\ref{ref_finale}). The experimental results are obtained for various particle radii $a=20\,\mu{\rm m},\,40\,\mu{\rm m},\,70\,\mu{\rm m},\,125\,\mu{\rm m},\,250\,\mu{\rm m}$ and viscosity of the interstitial fluid.}
\label{fig:final}
\end{figure}

In Fig. \ref{fig:final}, we have reported the experimental results for different fiber radii, particle radii, and viscosities. The results are expressed in terms of the threshold capillary number rescaled by the term containing the Goucher number dependence to emphasize that all the data are indeed described well by the simple power law dependence $(a/R)^{3/2}$.

The solid line in Fig. \ref{fig:final} is the power law on the right hand side of (\ref{ref_finale2}) with a ``best fit'' value of $\alpha\sim1.1$, which agrees well with the value obtained for a plate, $\alpha \simeq 1.1 \pm 0.1$ \cite[][]{sauret2019capillary}. This agreement suggests that the criterion of Eq. (\ref{eq:criterion}) holds despite moderately large particles, {\textit{i.e.} for particle to fiber aspect ratio $a/R$ of order 0.1}, and that the influence of the cylindrical geometry of the fiber is well characterized by our approach. In particular, the limit of Eq. (\ref{ref_finale}) as ${\rm Go}\to \infty$ reduces after some simplification to ${\rm Ca}^* \simeq 0.26\,(a/\ell_c)^{1/2}$, which is in agreement with previous experiments performed with a flat plate \cite[][]{sauret2019capillary}. {\cite{white1966theory} emphasized that the large cylinders or flat plates regime is obtained for Goucher number ${\rm Go} > 3$. According to equation (\ref{Alex_thickness}), $\mathrm{Go}>3$ means that the azimuthal curvature contributes less than 5\% deviation from the planar curvature. Therefore the expression proposed in this manuscript should be used for fibers of radius $R<3\,\ell_c$, which corresponds to $R \sim 4.2\,{\rm mm}$ for the fluids considered here}.

We note finally that the full finite film thickness expression of Eq. (\ref{system1}) can also be inverted for an explicit form in threshold capillary number in cases where the thin film approximation is not appropriate.


\section{Conclusion}

In this paper, we have characterized the entrainment of particles when withdrawing a fiber from a dilute suspension. We have shown that particles remain trapped in the liquid bath at small capillary numbers due to the capillary force exerted by the meniscus. At larger capillary numbers, individual particles can flow through the stagnation point and are entrained in the liquid film. We demonstrated that there is a strong dependence between this threshold capillary number and the fiber radius.

The results are rationalized by calculating the thickness of the film at the stagnation point for fibers. This thickness is related to the thickness of the liquid film through a prefactor $k=3$. Using this result, and the thickness of the coating film for various Goucher number, the experimental data can be rescaled. We can, therefore, predict whether particles will be entrained in the coating film depending on the Goucher number ${\rm Go}=R/\ell_c$ and the capillary number ${\rm Ca}$. Unlike the 2D situation with a plate, in which only the withdrawal rate affects the coating thickness and particle entrainment, the fiber radius provides a new level of control and an additional design criterion for dip coating applications, in particular for size-based particle sorting \cite[][]{dincau2019capillary}.

\medskip

\noindent \textbf{Acknowledgement.} This material is based upon work supported by the National Science Foundation under NSF Faculty Early Career Development (CAREER) Program Award CBET No. 1944844. We thank Bud Homsy, Carlo Colosqui and Howard Stone for helpful discussions and comments.

\medskip

\noindent \textbf{Declaration of Interests.} The authors report no conflict of interest.

\bibliographystyle{jfm}
\bibliography{BiblioDipCoating}

\end{document}